\documentstyle[prb,multicol,aps,epsfig]{revtex}
\renewcommand{\widetext}{\end{multicols}\global\columnwidth42.5pc}
\begin{document}

\newcommand{\be}{\begin{equation}}
\newcommand{\ee}{\end{equation}}
\newcommand{\bea}{\begin{eqnarray}}
\newcommand{\eea}{\end{eqnarray}}
\newcommand{\br}{{\bf r}}
\newcommand{\bk}{{\bf k}}
\newcommand{\bq}{{\bf q}}
\newcommand{\bn}{{\bf n}}
\newcommand{\bp}{{\bf p}}
\newcommand{\bE}{{\bf E}}
\newcommand{\E}{{\bf E}}
\newcommand{\ve}{\varepsilon}
\newcommand{\R}{{\bf R}}
\newcommand{\w}{\omega}
\newcommand{\St}{{\rm St}}
\newcommand{\nn}{{\bf i}}
\renewcommand{\j}{{\bf j}}
\renewcommand{\r}{{\bf r}}
\newcommand{\bzeta}{\mbox{\boldmath $\zeta$}}

\draft
\title{Theory of microwave-induced oscillations\\ in the 
magnetoconductivity of a 2D electron gas}
\author{I.A.~Dmitriev$^{1,*}$, M.G.~Vavilov$^{2}$, I.L.~Aleiner$^{3}$,
A.D.~Mirlin$^{1,4,\dagger}$, and D.G.~Polyakov$^{1,*}$}
\address{$^1$Institut f\"ur Nanotechnologie, Forschungszentrum
Karlsruhe, 76021 Karlsruhe, Germany}
\address{$^2$ 
Center for Materials Sciences and
Engineering,
Massachusetts Institute of Technology,
\\Cambridge, MA 02139, USA}
\address{$^3$Physics Department, Columbia University, New York, NY
10027, USA}
\address{$^4$Institut f\"ur Theorie der Kondensierten Materie,
Universit\"at Karlsruhe, 76128 Karlsruhe, Germany}
\maketitle
\begin{abstract}
We develop a theory of magnetooscillations in the photoconductivity of a
two-dimensional electron gas observed in recent experiments.  The effect is
governed by a change of the electron distribution function induced by the
microwave radiation. We analyze a nonlinearity with respect to both the {\it
dc} field and the microwave power, as well as the temperature dependence
determined by the inelastic relaxation rate.
\end{abstract}
\pacs{PACS numbers: 73.40.-c, 78.67.-n, 73.43.-f, 76.40.+b}
\begin{multicols}{2}
\narrowtext

\section{Introduction}
\label{s0}

Recent experiments have discovered \cite{zudov01} that the resistivity of a
high-mobility two-dimensional electron gas (2DEG) in GaAs/AlGaAs
heterostructures subjected to microwave radiation of frequency $\omega$
exhibits magnetooscillations governed by the ratio $\omega/\omega_c$, where
$\omega_c$ is the cyclotron frequency.  Subsequent work
\cite{mani02,zudov03,mani03,yang03,dorozhkin03,willett03} has shown that for
samples with a very high mobility and for high radiation power the minima of
the oscillations evolve into zero-resistance states (ZRS).

These spectacular observations have attracted much theoretical interest.  As
was shown in Ref.~\onlinecite{andreev03}, the ZRS can be understood as a
direct consequence of the oscillatory photoconductivity (OPC), provided that
the latter may become negative.  A negative value of the OPC signifies an
instability leading to the formation of spontaneous-current domains showing
zero value of the observable resistance. Therefore, the identification of the
microscopic mechanism of the OPC appears to be the key question in the
interpretation of the
data.\cite{zudov01,mani02,zudov03,mani03,yang03,dorozhkin03,willett03}

A mechanism of the OPC proposed in Ref.~\onlinecite{durst03} is based on the
effect of microwave radiation on electron scattering by impurities in a strong
magnetic field (see also Ref.~\onlinecite{ryzhii} for an earlier theory and
Ref.~\onlinecite{vavilov03} for a systematic theory).  An alternative
mechanism of the OPC was recently proposed in Ref.~\onlinecite{dmitriev03}. In
contrast to Refs.~\onlinecite {durst03,ryzhii,vavilov03}, this mechanism is
governed by a radiation-induced change of the electron distribution
function. Because of the oscillations of the density of states (DOS),
$\nu(\ve)$, related to the Landau quantization, the correction to the
distribution function acquires an oscillatory structure as well. This
generates a contribution to the {\it dc} conductivity which oscillates with
varying $\omega/\omega_c$. A distinctive feature of the contribution of
Ref.~\onlinecite{dmitriev03} is that it is proportional to the inelastic
relaxation time $\tau_{\rm in}$.  A comparison of the results of
Refs.~\onlinecite{vavilov03} and \onlinecite{dmitriev03} shows that the latter
contribution dominates if $\tau_{\rm in}\gg \tau_{\rm q}$ (where $\tau_{\rm
q}$ is the quantum, or single-particle, relaxation time due to impurity
scattering), which is the case for the experimentally relevant temperatures.

The consideration of Ref.~\onlinecite{dmitriev03} is restricted to the regime
which is linear in both the {\it ac} power and the {\it dc} electric
field. The purpose of this paper is to develop a complete theory of the OPC
governed by this mechanism, including nonlinear effects. We will demonstrate
that the conductivity at a minimum becomes negative for a large microwave
power and that a positive sign is restored for a strong {\it dc} bias, as it
was assumed in Ref.~\onlinecite{andreev03}.

The paper is organized as follows. First, in Sec.~\ref{s1} we formulate a
general approach to the problem. In Sec.~\ref{s2} we calculate the
non-equilibrium distribution function for overlapping Landau levels (LLs). In
Sec.~\ref{s3} we consider the OPC in the linear regime with respect to the
{\it dc} field. In Sec.~\ref{s4} we analyze the ZRS and calculate the
spontaneous electric field in the domains. In Sec.~\ref{s5} we turn to
separated LLs. Section \ref{s6} deals with the inelastic relaxation due to
electron-electron scattering. Finally, in Sec.~\ref{s7} we briefly discuss the
magnetooscillations in the Hall photoresistivity. In Sec.~\ref{s8} we
summarize our results and compare them with the experimental data. A brief
account of the results of this paper was presented in Ref.~\onlinecite{short}.

\section{General formalism}
\label{s1}

We consider a 2DEG (mass $m$, density $n_e$, Fermi velocity $v_F$) subjected
to a transverse magnetic field $B=(mc/e)\,\omega_c$. We assume that the field
is classically strong, $\omega_c\tau_{\rm tr}\gg 1$, where $\tau_{\rm tr}$ is
the transport relaxation time at $B=0$. The photoconductivity $\sigma_{\rm
ph}$ determines the longitudinal current flowing in response to a {\it dc}
electric field ${\cal E}_{\rm dc}$, $\vec{j}\cdot\vec{\cal E}_{\rm
dc}=\sigma_{\rm ph}{\cal E}_{\rm dc}^2$, in the presence of a microwave
electric field ${\bf{\cal E}}_\omega\cos\omega t$. The more frequently
measured \cite{zudov01,mani02,zudov03,mani03,dorozhkin03,willett03}
longitudinal resistivity, $\rho_{\rm ph}$, is given by $\rho_{\rm ph}\simeq
\rho_{xy}^2\sigma_{\rm ph}$, where $\rho_{xy}\simeq eB/n_ec$ is the Hall
resistivity, affected only weakly by the radiation.

We start with the formula for the {\it dc}  conductivity per spin:
\be
\label{photo}
\sigma_{\rm ph}=\int d\ve \,
\sigma_{\rm dc}(\ve)\, \left[-\partial_\ve
f(\ve)\right],
\ee
where $f(\ve)$ is the electron distribution function, and $\sigma_{\rm
dc}(\ve)$ determines the contribution of electrons with energy $\ve$ to the
dissipative transport. In the leading approximation,
\cite{vavilov03,dmitriev03}
\be
\sigma_{\rm dc}(\ve)=\frac{e^2\nu(\ve) v_F^2}{2}\;
\frac{\tau_{\rm tr,B}^{-1}(\ve)}{\omega_c^2+\tau_{\rm tr,B}^{-2}(\ve)}, 
\label{dc_long}
\ee
where $\tau_{\rm tr,B}$ is the transport scattering time in a quantizing
magnetic field, $\tau_{\rm tr,B}(\ve)=\tau_{\rm tr}\nu_0/\nu(\ve)$, and
$\nu_0=m/2\pi$ is the DOS per spin at zero $B$ (we use $\hbar=1$). We note
that Eq.~(\ref{dc_long}) has a Drude form with the DOS $\nu(\ve)$ and the
transport time $\tau_{\rm tr,B}(\ve)$ dependent of energy due to the Landau
quantization.

For a classically strong magnetic field, $\omega_c\tau_{\rm tr}\gg 1$, the
above expression reduces to
\be
\sigma_{\rm dc}(\ve)=\sigma^{\rm
D}_{\rm dc}\,\tilde{\nu}^2(\ve), 
\label{dc_short}
\ee
where $\sigma^{\rm D}_{\rm dc}=e^2\nu_0 v_{\rm F}^2/2\omega_c^2\tau_{\rm tr}$
is the {\it dc} Drude conductivity per spin in strong $B$ and we introduced
the dimensionless DOS, $\tilde{\nu}(\ve)=\nu(\ve)/\nu_0$.

We neglect here the effect of the microwaves on the impurity collision
integral, which yields a subleading contribution to the phoconductivity, as
discussed in Sec.~\ref{s3}.  The dominant effect is due to a non-trivial
energy dependence of the non-equilibrium distribution function $f(\ve)$.  The
latter is found as a solution of the stationary kinetic equation for the zero
angular harmonic of the distribution function $f(\ve)$:
\be
\label{kineq_gen}
\St_\omega\{f\}+\St_{\rm dc}\{f\}=-\St_{\rm in}\{f\}\,.
\ee
Here the left--hand side represents the effect of the microwaves ($\St_\omega
$) and of the $\it dc$ field ($\St_{\rm dc}$) in the presence of impurities
while the right--hand side accounts for the inelastic relaxation.

The first term on the left--hand side describes the absorption and emission of
microwave quanta; the rate of these transitions was calculated in 
Ref.~\onlinecite{dmitriev03}, yielding 
\bea\nonumber
\St_\omega\{f\}&=&\frac{{\cal E}^2_\omega}{4\omega^2}\sum_{\pm}
\frac{e^2 v_F^2 \,\tau_{\rm tr,B}^{-1}(\ve+\omega)\,[\,f(\ve+\omega)-f(\ve)\,]}
{2(\omega\pm\omega_c)^2+
\tau_{\rm tr,B}^{-2}(\ve+\omega)+\tau_{\rm tr,B}^{-2}(\ve)}\\
 &+& \{\omega\to-\,\omega\}\,.
\label{ac_long}
\eea
We will assume that $|\omega\pm\omega_c|\gg \tau_{{\rm tr,B}}^{-1}$, thus
excluding a narrow vicinity of the cyclotron resonance. This allows us to
neglect the $\ve$--dependent terms in the denominator, which reduces
Eq.~(\ref{ac_long}) to the form 
\be 
\St_\omega\{f\}={\cal
E}^2_\omega\,\frac{\sigma^{\rm D}_\omega}{2\omega^2\nu_0}
\sum\limits_{\pm}\tilde{\nu}(\ve\pm\omega)\,[\,f(\ve\pm\omega)-f(\ve)\,]\,,
\label{ac_short}
\ee
where the {\it ac} Drude conductivity per spin is given by
\be
\label{Drude}
\sigma^{\rm D}_\omega=\sum_{\pm}\,\frac{e^2\nu_0 v_{\rm F}^2}
{4 \tau_{\rm tr}(\omega\pm\omega_c)^2}.
\ee
Furthermore, we assume, in accordance with the experiments, a linear
polarization of the microwaves. For a circular polarization, one should retain
only one term on the right-hand side of Eq.~(\ref{Drude}), which, away from
the cyclotron resonance, does not affect the results in any essential way.

The second term on the left--hand side of Eq.~(\ref{kineq_gen}) represents the
effect of the {\it dc} electric field. The impurity scattering in a quantizing
magnetic field leads to the spatial diffusion with a diffusion coefficient
$D_{\rm B}(\ve)=v_F^2/2\omega_c^2\tau_{\rm tr,B}(\ve)$. In view of
conservation of the total energy $\ve+e{\cal E}_{\rm dc}x$ in a {\it dc} field
(directed along the $x$ axis) the spatial diffusion is translated into the
diffusion in $\ve$-space, with a diffusion coefficient $(e{\cal E}_{\rm dc})^2
D_{\rm B}(\ve)$ and the DOS $\nu(\ve)$,
\bea
\label{St_dc}
\nonumber
\St_{\rm dc}\{f\}&=&\nu^{-1}(\ve)\,\partial_\ve
\left[\,\nu(\ve)\,e^2{\cal E}_{\rm dc}^2D_{\rm B}(\ve)\,\partial_\ve\,f(\ve)\,
\right]\, \\
&=&{\cal E}^2_{\rm dc}\,\frac{\sigma^{\rm D}_{\rm 
dc}}{\nu_0\tilde{\nu}(\ve)} \,\partial_\ve\!
\left[\,\tilde{\nu}^2(\ve)\,\partial_\ve f(\ve)\,\right].
\eea
Equation (\ref{St_dc}) can also be obtained from Eq.~(\ref{ac_short}) by
taking the limit $\omega\to 0$ and replacing the period-average of the {\it
ac} field squared, ${1\over 2}{\cal E}_\omega^2$, by ${\cal E}_{\rm dc}^2$.

Substituting Eqs.~(\ref{ac_short}),(\ref{St_dc}) in the kinetic equation
(\ref{kineq_gen}), we get
\bea
\label{kineq}
\nonumber &&{\cal E}^2_\omega\,\frac{\sigma^{\rm D}_\omega}{2\omega^2\nu_0}
\sum\limits_{\pm}\tilde{\nu}(\ve\pm\omega)\,[\,f(\ve\pm\omega)-f(\ve)\,]\\
&&+\,\,{\cal E}^2_{\rm dc}\,\frac{\sigma^{\rm D}_{\rm
dc}}{\nu_0\tilde{\nu}(\ve)} \, \partial_\ve
\!\left[\,\tilde{\nu}^2(\ve)\,\partial_\ve f(\ve)\,\right]
={f(\ve)-f_T(\ve)\over\tau_{\rm in}}, 
\eea 
where the inelastic processes are included in the relaxation time
approximation and $f_T(\ve)$ is the Fermi distribution. A detailed discussion
of the inelastic relaxation and a calculation of the inelastic relaxation time
$\tau_{\rm in}$ are relegated to Sec.~\ref{s6}.

Equation (\ref{kineq}) suggests convenient dimensionless units for the
strength of the {\it ac} and {\it dc} fields:
\begin{mathletters}
\label{unitsPQ}
\bea
&&{\cal P}_\omega
=\frac{\tau_{\rm in}}{\tau_{\rm tr}}
\left(\frac{e {\cal E}_\omega v_F}{\omega}\right)^2
\frac{\omega_c^2+\omega^2}{(\omega^2-\omega_c^2)^2},
\label{units1}\\
&&{\cal Q}_{\rm dc}=\frac{2\,\tau_{\rm in}}{\tau_{\rm tr}}
\left(\frac{e {\cal E}_{\rm dc} v_F}{\omega_c}\right)^2
\left(\frac{\pi}{\omega_c}\right)^2.\label{units}
\eea
\end{mathletters}
With these notations, Eq.~(\ref{kineq}) reads
\bea\nonumber
&&\frac{{\cal P}_\omega}{4}
\sum\limits_{\pm}\tilde{\nu}(\ve\pm\omega)\,[\,f(\ve\pm\omega)-f(\ve)\,]\\
&&+{{\cal Q}_{\rm dc}\,\omega_c^2\over4\pi^2\tilde\nu(\ve)}\, \partial_\ve
\!\left[\,\tilde{\nu}^2(\ve)\,\partial_\ve f(\ve)\,\right]=f(\ve)-f_T(\ve)\,.
\label{kineq1}\eea
Note that ${\cal P}_\omega$ and ${\cal Q}_{\rm dc}$ are proportional to
$\tau_{\rm in}$ and are infinite in the absence of the inelastic relaxation
processes.

The left-hand side of the kinetic equation (\ref{kineq1}), as well as
Eqs.~(\ref{photo}),(\ref{dc_short}) for the photoconductivity, can also be
extracted from the quantum Boltzmann equation of Ref.~\onlinecite{vavilov03},
as we show in Appendix. Below, we use
Eqs.~(\ref{photo}),(\ref{dc_short}),(\ref{kineq1}) to analyze the
photoconductivity in the both limiting cases of overlapping and separated LLs.

\section{Non-equilibrium distribution function induced by microwave
  radiation}
\label{s2}

\begin{figure}[ht]
 \narrowtext
 \centerline{ {\epsfxsize=8cm{\epsfbox{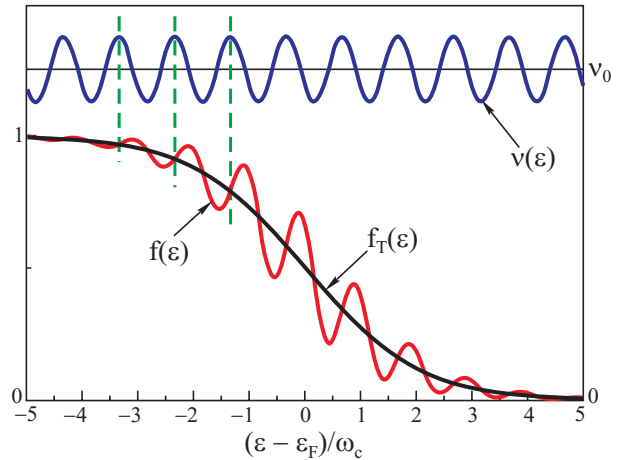}} }}
 \vspace{3mm}
 \caption{Schematic behavior of the oscillatory density of states $\nu(\ve)$ 
and radiation induced oscillations 
in the distribution function $f(\ve)$
for $\sin(2\pi \omega/\omega_c)\,>\,0$.}
 \label{fig0}
 \end{figure}

We consider first the case of overlapping LLs, with the DOS given by
\be
\label{delta}
\tilde{\nu}=1- 2\delta\cos
\frac{2\pi\ve}{\omega_c},\;\;\;
\delta=\exp\left(-\frac{\pi}{\omega_c\tau_{\rm q}}\right)\ll 1. 
\ee
Here $\tau_{\rm q}$ is the zero-$B$ single-particle relaxation time, which is
much shorter than the transport time in high-mobility structures, $\tau_{\rm
q}\ll \tau_{\rm tr}$ (because of the smooth character of a random potential of
remote donors). The existence of a small parameter $\delta$ simplifies
solution of the kinetic equation (\ref{kineq1}). To first order in $\delta$,
we look for a solution in the form
\be
\label{fe-expansion}
f=f_0+f_{\rm osc}+O(\delta^2), \ \ \ \ 
f_{\rm osc}\equiv \delta\,
{\rm Re}\left[f_1(\ve)\,e^{i\frac{2\pi\ve}{\omega_c}}\right].
\ee
We assume that the temperature (measured in energy units,
$k_{\rm B}\!=\!1$) is sufficiently high, $T\gg \omega_c$, implying a
scale separation between the smooth energy dependence of functions
$f_{0,1}(\ve)$ on a scale of the order of $T$ and the fast
oscillations with a period $\omega_c$~.\cite{note3} 
We also assume that the electric fields are not too strong [\,${\cal
P}_\omega(\omega/T)^2\ll 1$ and ${\cal Q}_{\rm dc}(\omega_c/T)^2\ll
1$\,], so that the smooth part $f_0(\ve)$ is close to the Fermi
distribution $f_T(\ve)$ at a bath temperature $T$;
otherwise, the temperature of the electron gas is further increased
due to heating.\cite{note3.1} We obtain\cite{note4}
\be
\label{distr}
f_{\rm osc}(\ve) = \delta\,\frac{\omega_c}{2\pi}\:
\frac{\partial f_T}{\partial \ve}\:
\sin \frac{2\pi\ve}{\omega_c}\:
\frac{{\cal P}_\omega\frac{2\pi \omega}{\omega_c}
\sin\frac{2\pi\omega}{\omega_c}
 +4{\cal Q}_{\rm dc}} {1+{\cal P}_\omega
 \sin^2\frac{\pi\omega}{\omega_c}
+{\cal Q}_{\rm dc}}.
\ee
Thus, the oscillations of the DOS $\nu(\ve)$ induce an oscillatory
contribution $f_{\rm osc}(\ve)$ to the distribution function, as illustrated
in Fig.~\ref{fig0}.
One might naively think that the small correction (\ref{distr}) to the
distribution function will only weakly affect the conductivity. This
is not the case, however. The reason is that, due to the fast oscillations
in $f_{\rm osc}$, the derivative $\partial_\ve f_{\rm osc}$ may be
large. As a result, a small variation of the distribution function
(\ref{distr}) can strongly affect the conductivity, Eq.~(\ref{photo}).
In particular, when the regions of an inverted population in $f(\ve)$
correspond to the maxima in $\nu(\ve)$ (as in Fig.~\ref{fig0}), the
linear--response conductivity may become negative, as we show below.

\section{Oscillatory photoconductivity}
\label{s3}

To calculate the photoconductivity, we substitute Eq.~(\ref{distr}) for
the distribution function into Eq.~(\ref{photo}). Performing the energy
integration in Eq.~(\ref{photo}), we assume (in conformity with the
experiment) that $T$ is much larger than the Dingle temperature, $T\gg
1/2\pi\tau_{q}$. Under this condition, the temperature smearing yields a
dominant damping factor of the Shubnikov-de Haas oscillations, $X/\sinh X$
with $X=2\pi^2 T/\omega_c$. The terms of order $\delta$ in Eq.~(\ref{photo})
are then exponentially suppressed,
$$
\delta \int
d\ve\, \cos \frac{2\pi\ve}{\omega_c}\,\partial_\ve f_T
\propto \delta\exp (-2\pi^2 T/\omega_c) \ll \delta^2,
$$
and can be neglected. The leading $\omega$-dependent contribution to
$\sigma_{\rm ph}$ comes from the $\delta^2$ term generated by the product of
$\partial_\ve f_{\rm osc}(\ve) \propto \delta\cos \frac{2\pi\ve}{\omega_c}$
and the oscillatory part $-2\delta \cos\frac{2\pi\ve}{\omega_c}$ of
$\tilde{\nu}(\ve)$. This term does survive the energy averaging, $-\int d\ve\,
\cos^2\frac{2\pi\ve}{\omega_c}\,\partial_\ve f_T \simeq 1/2$. We thus find
\be
\label{result}
\frac{\sigma_{\rm ph}\ }{\sigma^{\rm D}_{\rm dc}}=1+2\delta^2\left[\,1
-
\frac{{\cal P}_\omega\frac{2\pi \omega}{\omega_c}
\sin\frac{2\pi\omega}{\omega_c}
 +4{\cal Q}_{\rm dc}} {1+{\cal P}_\omega
 \sin^2\frac{\pi\omega}{\omega_c}
+{\cal
Q}_{\rm dc}}
\right].
\ee

Equation (\ref{result}) is our central result. It describes the
photoconductivity in the regime of overlapping LLs, including all non-linear
(in ${\cal E}_\omega$ and ${\cal E}_{\rm dc}$) effects.  Let us analyze it in
more detail. In the linear-response regime (${\cal E}_{\rm dc}\to 0$) and for
a not too strong microwave field, Eq.~(\ref{result}) yields a correction to
the dark {\it dc} conductivity $\sigma_{\rm dc}= \sigma^{\rm D}_{\rm
dc}(1+2\delta^2)$, which is linear in the microwave power:
\be
\label{linear}
{\sigma_{\rm ph}-\sigma_{\rm dc}\over\sigma_{\rm dc}} = - 4 \delta^2
{\cal P}_\omega \,{\pi\omega\over\omega_c} \,\sin
{2\pi\omega\over\omega_c},
\ee
in agreement with Ref.~\onlinecite{dmitriev03}.  It is enlightening to compare
Eq.~(\ref{linear}) with the contribution of the effect of the {\it ac} field
on the impurity scattering.  \cite{durst03,ryzhii,vavilov03} The analytic
result, Eq.~(6.11) of Ref.~\onlinecite{vavilov03}, in the notation of
Eq.~(\ref{unitsPQ}) is
\be
{\sigma_{\rm ph}^{[11]}-
\sigma_{\rm dc}\over\sigma_{\rm dc}} =
-  12 \frac{\tau_{{\rm q}} }{\tau_{\rm in}}
\delta^2 {\cal P}_\omega
\left({\pi\omega\over\omega_c} \,\sin
{2\pi\omega\over\omega_c} + \sin^2\frac{\pi\omega}{\omega_c}
\right).
\label{sigma-collision}
\ee
This result has a similar frequency dependence as Eq.~(\ref{linear}); however,
its amplitude is much smaller at $\tau_{\rm in} \gg \tau_{{\rm q}}$, i.e., the
mechanism of Refs.~\onlinecite{durst03,ryzhii,vavilov03} appears to be
irrelevant. Physically, the effect of the {\it ac} field on the distribution
function is dominant because it is accumulated during a diffusive process of
duration $\tau_{\rm in}$, whereas Refs.~\onlinecite{durst03,ryzhii,vavilov03}
consider only one scattering event. Apart from the magnitude, the two
contributions are qualitatively different in their temperature and
polarization dependence. Specifically, the contribution related to the change
of the distribution function is strongly temperature-dependent (due to the
$T$-dependence of $\tau_{\rm in}$, see Sec.~\ref{s6}) and does not depend on
the direction of the linear polarization of the microwave field.  On the other
hand, the effect of microwaves on the impurity collision integral yields a
$T$-independent contribution which depends essentially on the relative
orientation of the fields $\vec{\cal E}_\omega$ and $\vec{\cal E}_{\rm dc}$
[Eq.~(\ref{sigma-collision}) represents the result averaged over the
polarization direction].

\begin{figure}[ht]
 \narrowtext
 \centerline{ {\epsfxsize=8cm{\epsfbox{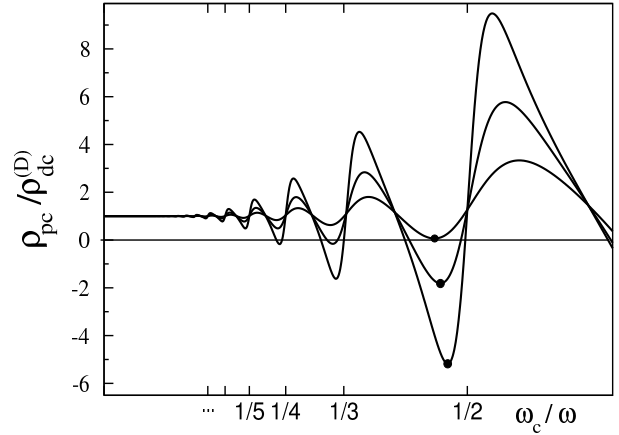}} }}
 \vspace{3mm}
 \caption{Photoresistivity (normalized to the dark Drude value) for
overlapping Landau levels vs $\omega_c/\omega$ at fixed
$\omega\tau_{\rm q}=2\pi$. The curves correspond to
different levels of microwave power ${\cal P}_\omega^{(0)}
=\{0.24,\,0.8,\,2.4\}$. Nonlinear $I-V$ characteristics at the marked
minima are shown in Fig.~\ref{fig2}.}
 \label{fig1}
 \end{figure}

With increasing microwave power, the photoconductivity saturates at the value
\be
\label{saturation}
{\sigma_{\rm ph}\over\sigma_{\rm dc}} = 1- 8 \delta^2
\,{\pi\omega\over\omega_c} \,\cot {\pi\omega\over\omega_c}~, \quad {\cal
P}_\omega\sin^2 {\pi\omega\over\omega_c}\gg 1.  
\ee 
Note that although the correction is proportional to $\delta^2\ll 1$, the
factor $8\pi(\omega/\omega_c)\cot(\pi\omega/\omega_c)$ is large in the
vicinity of the cyclotron resonance harmonics $\omega=k\omega_c$
($k=1,2,\ldots$), and allows the photo-induced correction to exceed in
magnitude the dark conductivity $\sigma_{\rm dc}$.  In particular,
$\sigma_{\rm ph}$ around minima becomes negative at ${\cal P}_\omega >{\cal
P}_\omega^* > 0 $, with the threshold value given according to
Eq.~(\ref{result}) by
\be
{\cal P}_\omega^*=
\left(4\delta^2\frac{\pi\omega}{\omega_c}\sin
\frac{2\pi\omega}{\omega_c}
-\sin^2\frac{\pi\omega}{\omega_c}\right)^{-1}.
\label{P*}
\ee
The evolution of the $B$ dependence of the photoresistivity $\rho_{\rm ph}$
with increasing microwave power is illustrated in Fig.~\ref{fig1}.  More
specifically, the curves in Fig.~\ref{fig1} correspond to different values of
the dimensionless parameter
\be
\label{P0}
{\cal P}_\omega^{(0)}\equiv{\cal P}_\omega(\omega_c=0)
=\frac{e^2}{\hbar c}\cdot\frac{\tau_{\rm in}}{\tau_{\rm tr}}
\cdot\frac{v_F^2}{\omega^2 S}\cdot\frac{8\pi P}{\hbar\omega^2},
\ee
where $P=c |{\cal E}_\omega|^2 S/8\pi$ is the microwave power over the sample
area $S$, $c$ is the speed of light, and we restored the Planck constant for
convenience.

\section{Zero--Resistance States}
\label{s4}

\begin{figure}
 \narrowtext
 \centerline{ {\epsfxsize=8cm{\epsfbox{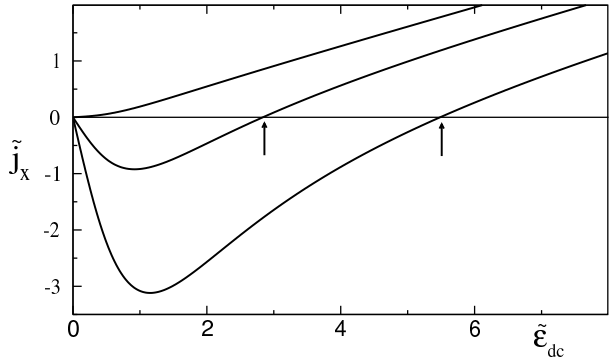}} }}
 \vspace{3mm}
 \caption{Current--voltage characteristics [\,dimensionless current
$\tilde{j}_x=(\sigma_{\rm ph}/\sigma^{\rm D}_{\rm dc}) \tilde{\cal
E}_{\rm dc}$ vs dimensionless field $\tilde{\cal E}_{\rm dc}={\cal
Q}_{\rm dc}^{1/2}$\,] at the points of minima marked by the circles in
Fig.~1. The arrows show the {\it dc} field $\tilde{\cal E}_{\rm dc}^*$
in spontaneously formed domains.}
\label{fig2}
 \end{figure}

Let us now fix $\omega/\omega_c$ such that ${\cal P}_\omega^*>0$, and consider
the dependence of $\sigma_{\rm ph}$ on the {\it dc} field ${\cal E}_{\rm dc}$
at a sufficiently strong microwave power ${\cal P}_\omega>{\cal P}_\omega^*$,
corresponding to the negative linear--response photoconductivity. As follows
from Eq.~(\ref{result}), in the limit of large ${\cal E}_{\rm dc}$ the
conductivity is close to the Drude value and thus positive, $\sigma_{\rm
ph}=(1-6\delta^2)\sigma_{\rm dc}^{\rm D} > 0$. Therefore, $\sigma_{\rm ph}$
changes sign at a certain value ${\cal E}_{\rm dc}^*$ of the {\it dc} field
(see Fig.~\ref{fig2}), which is determined by the condition ${\cal Q}_{\rm
dc}= ({\cal P}_\omega-{\cal P}_\omega^*)/{\cal P}_\omega^*$. The
negative-conductivity state at ${\cal E}_{\rm dc}<{\cal E}_{\rm dc}^*$ is
unstable with respect to the formation of domains with a spontaneous electric
field of the magnitude ${\cal E}_{\rm dc}^*$. \cite{andreev03} Using
Eqs.~(\ref{unitsPQ}),~(\ref{P*}), we obtain
\bea
\label{e-domain}\nonumber
{\cal E}_{\rm dc}^*&=&{1\over\sqrt{2}\,\pi}\,
{\omega_c\over e R_c}\left({\tau_{\rm
      tr}\over \tau_{\rm in}}\right)^{1/2}
\left[\left({{\cal E}_\omega\over
{\cal E}_\omega^*}\right)^2-1\right]^{1/2}
\\\nonumber &=&
\sqrt{{\cal E}_\omega^2-({\cal E}_\omega^*)^2}
\left[\frac{\omega_c^4 (\omega^2+\omega_c^2)}{2
\omega^2(\omega^2-\omega_c^2)^2}\right]^{1/2}\\
&\times&
{1\over\pi}{\rm Re}\left(4\delta^2\frac{\pi\omega}{\omega_c}\sin
\frac{2\pi\omega}{\omega_c}
-\sin^2\frac{\pi\omega}{\omega_c}\right)^{1/2},
\eea
with ${\cal E}_\omega^*$ being the threshold value of the {\it ac} field at
which the ZRS develops and $R_c=v_F/\omega_c$ the cyclotron radius.  Equation
(\ref{e-domain}) relates the electric field formed in the domains (measurable
by local probe \cite{willett03}) with the excess power of microwave
radiation. It is worth noticing that the last expression for ${\cal E}_{\rm
dc}^*$ does not explicitly contain the rate of the inelastic processes.

\section{Separated Landau levels}
\label{s5}

\begin{figure}
 \narrowtext
 \centerline{ {\epsfxsize=8cm{\epsfbox{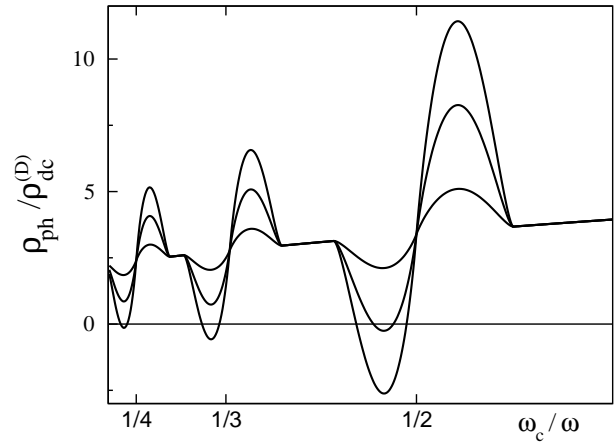}} }}
 \vspace{3mm}
 \caption{Photoresistivity (normalized to the Drude value) for
separated Landau levels vs $\omega_c/\omega$ at fixed
$\omega\tau_{\rm q}=16\pi$. The curves correspond to
different levels of microwave power ${\cal P}_\omega^{(0)}
=\{0.01,\,0.03,\,0.05\}$. }
 \label{fig3}
 \end{figure}

We now turn to the regime of strong $B$, $\omega_c\tau_{{\rm q}}/\pi\gg
1$, where the LLs get separated. The DOS is then given (within the
self-consistent Born approximation) by a sequence of semicircles of
width $2\Gamma=2(2\omega_c/\pi\tau_{{\rm q}})^{1/2}$:
\be
\label{SepDOS}
\tilde{\nu}(\ve)=\frac{2\omega_c}{\pi\Gamma^2}
\sum_n {\rm Re}\,
\sqrt{\Gamma^2-\left(\ve-n\omega_c-\omega_c/2\right)^2}.
\ee
We use Eqs.~(\ref{photo}) and (\ref{kineq1}) to evaluate the OPC at ${\cal
Q}_{dc}\to 0$ to first order in $\cal P_\omega$ and estimate the correction of
the second order. The condition $T\gg\omega_c$ allows us to separate the slow
dependence on $\ve$ on the scale of $T$ and fast oscillations with the period
$\omega_c$ in the integral (\ref{photo}) by averaging over the period of the
oscillations.  After integrating the resulting slow-varying functions we
obtain
\bea
\label{Sep_ph}\nonumber
&&{\sigma_{\rm ph}\over\sigma^{\rm D}_{\rm dc}}
=\langle\,\tilde{\nu}^2(\ve)\,\rangle_\ve\\\nonumber
&& -\,{\omega{\cal P}_\omega\over4}\,
\langle\,[\,\tilde{\nu}(\ve+\omega)-\tilde{\nu}(\ve-\omega)\,]
\,\partial_\ve\tilde{\nu}^2(\ve)\,\rangle_\ve\\\nonumber
&&+\,\omega{\cal P}^2_\omega\sum\nolimits_{k,l}\,a_{k,\,l}\,
\langle\,\tilde{\nu}(\ve+k\omega)\,\tilde{\nu}(\ve+l\omega)\,
\partial_\ve\tilde{\nu}^2(\ve)\,\rangle_\ve\\
&&+O(\,{\cal P}^3_\omega\,).
\eea
Here the angular brackets denote averaging over $\ve$ within the period
$\omega_c$, and $a_{k,\,l}$ are numerical coefficients. The result for
separated LLs, Eq.~(\ref{SepDOS}), reads
\bea 
\label{1order}
&&{\sigma_{\rm ph}\over\sigma^{\rm D}_{\rm dc}}
={16\omega_c\over3\pi^2\Gamma}  
\left\{1-{\cal P}_\omega {\omega\omega_c\over\Gamma^2} \right. \nonumber \\
&& \phantom{aaa} \times \left.\left[
\sum_n \Phi\left({\omega-n\omega_c\over\Gamma}\right)
+ O\left({\omega_c {\cal P}_\omega\over\Gamma}\right)
\right] \right\}, \\
&&\Phi(x)\!=\!\frac{3 x}{4\pi}{\rm Re}\!\left[ {\rm arccos}(|x|-1)-
{1+|x|\over 3}\sqrt{|x|(2-|x|)} \right].
\nonumber
\eea
The photoresistivity for the case of separated LLs, Eq.~(\ref{1order}), is
shown in Fig.~\ref{fig3} for several values ${\cal P}_\omega$ of the microwave
power.  Notice that a correction to Eq.~(\ref{1order}) of second order in
${\cal P}_\omega$ is still small even at microwave power exceeding the
threshold value
\be
{\cal  P}_\omega^*\sim\Gamma^2/\omega\omega_c, 
\ee
since $\omega_c {\cal P}_\omega^* /\Gamma \sim \Gamma/\omega \ll 1$.  This
means that it suffices to keep the linear-in-${\cal P}_\omega$ term only even
for the microwave power ${\cal P}_\omega>{\cal P}^*_\omega$ at which the
linear-response resistance becomes negative.

As in the case of overlapping LLs, a negative value of the linear-response
conductivity signals an instability leading to the formation of domains with
the field ${\cal E}^*_{\rm dc}$ at which $\sigma_{\rm ph}({\cal E_{\rm
dc}})=0$. It turns out, however, that for separated LLs the kinetic equation
in the form of Eq.~(\ref{kineq}) yields zero (rather than expected positive)
conductivity in the limit of strong ${\cal E}_{\rm dc}$.\cite{note1} This
happens because elastic impurity scattering between LLs, inclined in a strong
{\it dc} field, is not included in Eq.~(\ref{kineq}).  The inter-LL
transitions become efficient in {\it dc} fields as strong as [see Eq.~(5.5) of
Ref.~\onlinecite{vavilov03}]
\be
{\cal E}_{\rm dc}^*\sim \left({\tau_{\rm
tr}\over\tau_{\rm q}}\right)^{1/2}\,{\omega_c\over e R_c}\,,
\ee 
which actually gives the strength of the field in domains for the case
of separated LLs.

\section{Inelastic relaxation due to electron--electron collisions}
\label{s6}

Finally, we calculate the inelastic relaxation time $\tau_{\rm in}$.  Of
particular importance is its $T$ dependence which in turn determines that of
$\sigma_{\rm ph}$. At not too high $T$, the dominant mechanism of inelastic
scattering is due to electron-electron (e-e) collisions. It is worth
emphasizing that the e-e scattering does not yield relaxation of the total
energy of the 2DEG and as such cannot establish a steady-state {\it dc}
photoconductivity. That is to say the smearing of $f_0(\ve)$ in
Eq.~(\ref{fe-expansion}), which is a measure of the degree of heating, is
governed by electron-phonon scattering. However, the e-e scattering at
$T\gg\omega_c$ does lead to relaxation of the oscillatory term $f_{\rm osc}$,
Eq.~(\ref{distr}), and thus determines the $T$ behavior of the oscillatory
contribution to $\sigma_{\rm ph}$.

Quantitatively, the effect of electron-electron interaction is taken
into account by replacing the right-hand side of Eq.~(\ref{kineq})
by $-{\rm St}_{ee}\left\{ f \right\}$, where the collision integral
${\rm St}_{ee}\left\{ f \right\}$ is given by
\bea
\label{St}\nonumber
{\rm St}_{ee}\left\{ f \right\}&=&
\int d\ve^\prime\int dE \: A(E,\,\ve,\,\ve^\prime)\\\nonumber
&\times&\left[
-f(\ve)f^{(h)}(\ve_+)f(\ve^\prime) f^{(h)}(\ve^\prime_-)\right.\\
&&\phantom{lkjkjfh}\left.+  f^{(h)}(\ve)f(\ve_+)f^{(h)}(\ve^\prime) f(\ve^\prime_-)
\right],   
\eea
and $f^{(h)}(\ve) \equiv 1-f(\ve)$, $\ve_{+}= \ve + E$, $\ve_{-}^\prime=
\ve^\prime - E$.  The function $A(E,\,\ve,\,\ve^\prime)$ describes the
dependence of the matrix element of the screened Coulomb interaction on the
transferred energy $E$ and the energies of colliding particles. In what
follows we use the linearized form of ${\rm St}_{ee}$.

\subsection{Overlapping Landau levels}
\label{ss61}

For overlapping LLs, we put $\tilde\nu =1$ in accord with the accuracy of
Eq.~(\ref{result}).  Then the kernel in Eq.~(\ref{St}) depends on the
transferred energy $E$ only and is given by the general formula
\cite{altshuler85}
\be\label{A}
A(E)=\frac{2\nu_0}{\pi}\int\!\frac{d^2 q}{(2\pi)^2}\,\,
|U|^2\,({\rm Re}\,\langle {\cal D}\rangle)^2\,.
\ee
Here the factor of two accounts for spin, $\langle {\cal D}\rangle(E,q)$ is
the angle-averaged particle--hole propagator,
\be\label{U}
U(E,q)= \frac{\kappa/2\nu_0}{q+\kappa(1+iE\langle {\cal D}\rangle)}
\ee
the dynamically screened Coulomb potential, and $\kappa=4\pi e^2 \nu_0$ the
inverse screening length.

The propagator $\langle {\cal D}\rangle(E,q)$ has the (disorder-independent)
Fermi-liquid form for ultra--ballistic momenta, $q\gg q_1=(\omega_c/D)^{1/2}$,
where $D=R_c^2/2\tau_{\rm tr}$ is the diffusion coefficient in a classically
strong magnetic field,
\be\label{D_FL}
\langle{\cal D}\rangle= (\,q^2 v_F^2-E^2\,)^{-1/2}~,\quad q\gg q_1=
(\omega_c/D)^{1/2}~.
\ee
With lowering $q$ it crosses over into the quasiclassical particle--hole
propagator in a smooth random potential (``ballistic diffuson'')\cite{RAG,GM}
\be\label{D}
\langle{\cal D}\rangle= \sum\limits_{n=0}^\infty\;
\frac{J^2_n(q R_c)}{-i(E-n\omega_c)+D q^2+n^2/\tau_{\rm tr}},\quad q\ll q_1~,
\ee
where $J_n(x)$ is the Bessel function.

Let us calculate the kernel $A(E)$ for $E\lesssim T$ within the
(experimentally relevant) temperature range,
\be
\omega_c\ll T\ll \omega_c (\omega_c\tau_{\rm tr})^{1/2}.
\label{Tcond1}
\ee
The integration domain in Eq.~(\ref{A}) naturally divides into two parts: $q>
q_1$ and $q< q_1$.  In view of Eq.(\ref{Tcond1}) the propagator (\ref{D_FL})
for ultra--ballistic momenta $q\gg q_1$ does not depend on the transferred
energy $E\lesssim T$; specifically, $\langle {\cal D}\rangle\simeq 1/qv_F$,
while the screening is effectively static, $U(E,q)\simeq U(0,q)$. It follows
that the contribution $A^>(E)$ of $q\agt q_1$ to the integral (\ref{A}) has
the form
\be\label{A>}
A^>(E)={1\over2\pi\ve_F}\ln\frac{\kappa}{q_1}.
\ee
By contrast, the contribution of $q\alt q_1$ to Eq.~(\ref{A}) 
is strongly $E$-dependent and is found as a sum over peaks coming
from different $n$ in Eq.~(\ref{D}), 
\be
\label{A<}
A^<(E)=\sum\limits_n A_n^<(E).
\ee

When calculating the contribution of a single peak $A_n^<$, the
following approximations are justified: in Eq.~(\ref{D}), the product
of the Bessel functions $J^2_n(q R_c)$ for the relevant momenta $q\gg
R_c^{-1}$ can be replaced by $2\cos^2\varphi_n/\pi q R_c$ with
$\varphi_n=q R_c+\pi(2 n+1)/4$; the term $n^2/\tau_{\rm tr}$ in
Eq.~(\ref{D}) can be neglected at $E\lesssim T$ in view of
Eq.~(\ref{Tcond1}); also, the first term in the denominator of
Eq.~(\ref{U}) can be omitted for $q\alt q_1\ll\kappa$.  Introducing the
dimensionless parameters $\Delta_n(E)=|E/\omega_c-n|\leq 1/2$,
$\beta_n=2\pi^2\omega_c\tau_{\rm tr}/n^2\gg 1$, and
$\gamma_n(E)=\beta_n^{1/3}\Delta_n(E)$ we express the contribution of
the $n$th peak for $\Delta_n\ll 1$ as\cite{crossover}
\bea
\nonumber
&&A_n^<(E)=\frac{\beta_n^{2/3}}{4\pi^3\ve_F}\int\nolimits^{\infty}_0
\!\!\!{\rm d}x
\frac{x^5}{x^4+\gamma_n^2}\\
&&\times\left[\frac{x^6}{4\cos^4(\alpha_n x)}
+\left(1-\frac{\gamma_n x}{2\cos^2(\alpha_n x)}\right)^2\right]^{-1}.
\label{integral}
\eea
After averaging over the fast oscillations with $\alpha_n x$, where
$\alpha_n=n\beta_n^{1/3}/\pi$, the integration yields 
\be\label{A<_n}
A_n^<(E)={1\over4\pi^2\ve_F}
\left\{
\matrix{c\;\beta_n^{2/3}, & \Delta_n\ll\beta_n^{-1/3}\cr
3/8\pi\Delta_n^{2}, & 1\gg \Delta_n\gg\beta_n^{-1/3}
}
\right.\,,
\ee
where $c=\Gamma(7/6)/2^{1/3}(3\pi)^{1/2}\Gamma(2/3)\simeq 0.18$.  We now use
Eqs.~(\ref{St}),~(\ref{A>})--(\ref{A<_n}) to calculate the relaxation rate
$\tau^{-1}_{\rm in}$ of the oscillatory component of the distribution
function, $f_{\rm osc}(\ve)$.  As we are going to show, it is dominated by the
large--momentum transfers, $q\gg q_1$. The contribution of this
large--momentum region is easily evaluated: since $A^>(E)$ is
energy-independent, it is sufficient to take into account the out-scattering
term only.  We thus return to the right-hand side of Eq.~(\ref{kineq}) with
the inelastic relaxation rate $\tau^{-1}_{\rm in}$ replaced by\cite{note2}
\bea\label{out}\nonumber
\tau_{ee}^{-1}&=&\int\limits_{-\infty}^{\infty}\!dE\;A^>(E)
\;\frac{E}{2}\,\left(\coth\frac{E}{2T}-\tanh\frac{E+\ve}{2T}\right)\\\label{overLL}
&=&
\frac{\pi^2 T^2+\ve^2}{4 \pi\ve_F}\ln\frac{\kappa v_F}
{\omega_c(\omega_c\tau_{\rm tr})^{1/2}}~.
\eea
Note that $\tau_{ee}^{-1}$ in Eq.~(\ref{overLL}) depends on $B$ through the
logarithmic factor only and crosses over into the conventional zero-$B$ result
$\tau_{ee}^{-1}=(\pi T^2/4\ve_F)\ln(\kappa v_F/T)$ when $T$ exceeds
$\omega_c(\omega_c\tau_{\rm tr})^{1/2}$.

Let us now turn to the contribution of the region $q\alt q_1$.  Evaluating the
out-scattering term [similar to the first line of Eq.~(\ref{out}), with
$A^>(E)$ replaced by $A^<(E)$, Eqs.~(\ref{A<}),(\ref{A<_n})], we find
$(\tau_{\rm ee}^<)^{-1}\sim(\omega_c\tau_{\rm tr})^{1/3}\,T^2/\ve_F$, which
can exceed the large--$q$ contribution (\ref{out}).  However, the relaxation
time approximation is no longer valid for $q<q_1$. Indeed, the main
contribution to $(\tau^<_{\rm ee})^{-1}$ comes from the energy transfers $E$
close to $n\,\omega_c$, $\Delta_n(E)\sim\beta_n^{-1/3}$, see
Eq.~(\ref{A<}). Such processes are inefficient as far as the relaxation of
$f_{\rm osc}$ is concerned, since the energy transfer is almost commensurate
with the period of the oscillations in the distribution function.  In other
words, the out--scattering term is almost compensated by the in-scattering
one, so that $A^<(E)=\sum_n A_n^<(E)$ is effectively replaced by $\sum_n
\Delta_n^2 A_n^<(E)$.  As a result, the contribution of region of $q\alt q_1$
to the relaxation rate is dominated by $q\sim q_1$ and is given by
Eq.~(\ref{out}) without the logarithmic factor, and thus can be neglected.

In fact, the situation is similar to the momentum relaxation due to
small-angle scattering off a smooth random potential. The momentum relaxation
time $\tau_{\rm tr}$ and the out-scattering (or single-particle) relaxation
time $\tau_{\rm q}$ in that case differ by the factor $(1-\cos\phi)$, which
accounts for a reduced contribution to the resistivity of small-angle
scattering with $\phi\ll 1$:
\bea\nonumber
\left.
\begin{array}{l}
\tau_{\rm q}^{-1} \\
\tau_{\rm tr}^{-1}
\end{array}
\right\}
= 2\pi\nu_0\int{d\phi\over 2\pi} \,W(2k_F\sin{\phi\over 2})\times
\left\{
\begin{array}{l}
1\\
(1-\cos\phi)
\end{array} \right.\,,
\eea
where $W(q)$ is the Fourier transform of the correlation function of the
random potential. A similar result for the relaxation of the oscillatory part
of the distribution functions is obtained from Eq.~(\ref{St}).  We linearize
Eq.~(\ref{St}) with the distribution function in the form (we assume
$T\gg\omega$)
\be
 f=f_T+\varphi(\ve)\;\partial_\ve f_T~,
\label{phi}
\ee 
where $\varphi(\ve)=\varphi(\ve+\omega_c)$. This ansatz is suggested
by Eq.~(\ref{distr}) and will be confirmed by the calculation
below. We substitute Eq.~(\ref{phi}) in Eq.~(\ref{St}) and use the
condition $T\gg\omega_c$ which allows us to separate the slow
dependence on $E,\,\ve^\prime$ on the scale of $T$ and fast
oscillations with the period $\omega_c$ by averaging over the period
of the oscillations. Using $A(E)=A(E+\omega_c)$ [see Eqs.~(\ref{A>}),
(\ref{A<}), and (\ref{A<_n})], we obtain the following integrals 
over the slow variables which all produce the same result,
\bea
\label{int}\nonumber
&&-\int d\ve^\prime\int dE \: [f_T^{(h)}(\ve_+)f_T(\ve^\prime) f_T^{(h)}(\ve^\prime_-)\\\nonumber
&&\phantom{qurtypqirutqyqpieru}
+f_T(\ve_+)f_T^{(h)}(\ve^\prime) f_T(\ve^\prime_-)]\,\partial_\ve f_T(\ve)\\\nonumber
&&=\int d\ve^\prime\int dE \: [f_T(\ve)f_T(\ve^\prime) f_T^{(h)}(\ve^\prime_-)\\\nonumber
&&\phantom{qurtypqirutqyqpieru}
+f_T^{(h)}(\ve)f_T^{(h)}(\ve^\prime) f_T(\ve^\prime_-)]\,\partial_\ve f_T(\ve_+)\\\nonumber
&&=-\int d\ve^\prime\int dE \: [f_T(\ve)f_T^{(h)}(\ve_+)f_T^{(h)}(\ve^\prime_-)\\\nonumber
&&\phantom{qurtypqirutqyqpieru}
+f_T^{(h)}(\ve)f_T(\ve_+)f_T(\ve^\prime_-)]\,\partial_{\ve^\prime} f_T(\ve^\prime)\\\nonumber
&&=\int d\ve^\prime\int dE \: [f_T(\ve)f_T^{(h)}(\ve_+)f_T(\ve^\prime)\\\nonumber
&&\phantom{qurtypqirutqyqpieru}
+f_T^{(h)}(\ve)f_T(\ve_+)f_T^{(h)}(\ve^\prime)]\,\partial_{\ve^\prime}f_T(\ve^\prime_-)\\
&&=\frac{\pi^2 T^2+\ve^2}{2}\partial_\ve f_T(\ve)\,.   
\eea
The collision integral thus reads
\bea
\label{St_tr}
&&-{\rm St}_{\rm in}\left\{ f \right\}=
\frac{\pi^2 T^2+\ve^2}{2}\:\frac{\partial f_T}{\partial \ve}\\ \nonumber
&&\times\langle\,A(E)\,[\,\varphi(\ve)-\varphi(\ve+E)
+\varphi(\ve^\prime)-\varphi(\ve^\prime-E)\,]\,\rangle_{\varepsilon^{\prime},\,
E}\,\,,
\eea
where the angular brackets denote
averaging over $\ve^\prime$ and $E$ within the period $\omega_c$. For a
harmonic modulation of the distribution function, 
$\varphi(\ve)\propto \cos(2\pi\ve/\omega_c+\theta)$, as in
Eq.~(\ref{fe-expansion}), and using $A(E)=A(-E)$, we obtain
\bea
\label{tau_in}\nonumber
&&-{\rm St}_{\rm in}\left\{ f\right\}=\frac{f-f_T}{\tau_{\rm in}}\;,\\
&&\tau_{\rm in}^{-1}=\frac{\pi^2 T^2+\ve^2}{2}\,
\langle\,A(E)\,[\,1-\cos(2\pi E/\omega_c)\,]\,\rangle_E\,.
\eea
Because of the factor $1-\cos(2\pi E/\omega_c)$ the contribution of small
momenta (\ref{A<}), (\ref{A<_n}) to the relaxation rate is small compared to
that of $q\agt q_1$. In the latter case, due to the energy-independent kernel
$A(E)$, Eq.~(\ref{A>}), the in-scattering part is zero on average, and
$\tau_{\rm in}^{-1}$ coincides with the out-scattering rate (\ref{out}).

The inelastic relaxation time $\tau_{\rm in}$ as obtained above
[Eqs.~(\ref{out}),(\ref{tau_in})] depends on energy $\ve$.\cite{conserve} This
makes the problem somewhat more complicated than the model considered in
Sec.~\ref{s1} with a phenomenological, $\ve$--independent parameter $\tau_{\rm
in}$. However, characteristic energies are $\ve\sim T$ [see Eq.~(\ref{phi})],
so that the $\ve$--dependence in Eqs.~(\ref{out}),(\ref{tau_in}) does not
change the $T^{-2}$ scaling of $\tau_{\rm in}$ but only yields a numerical
factor.  In particular, repeating the analysis of Eq.~(\ref{kineq}) in the
linear--in--${\cal P}_\omega$ regime with the $\ve$--dependent $\tau_{\rm
in}$, we find that $\tau_{\rm in}$ entering Eq.~(\ref{units1}) is effectively
replaced by
\bea
\nonumber
&&\int\! d\ve\,\tau_{\rm in}(\ve,T)\,(-\,\partial_\ve f_T)\\
&&{\ \ \ \ }=\tau_{\rm in}(0,T)\int\! d\ve\frac{-\,\partial_\ve
  f_T}{1+(\ve/\pi T)^2} 
\simeq0.822\,\tau_{\rm in}(0,T).
\label{tau>}
\eea
Therefore, at $T\gg\omega$ the linear photoresistivity, Eq.~(\ref{linear}),
scales as $T^{-2}$.

At $T\ll\omega$ (but still $T\gg\omega_c$), the $\ve$--dependence in
$\tau_{\rm in}$ becomes more important. Indeed, solving Eq.~(\ref{kineq}) to
the linear order in ${\cal E}_\omega^2$ (and at ${\cal E}_{\rm dc}=0$), we
find the following oscillatory contribution to the distribution function
\bea\nonumber
f_{\rm osc}(\ve)&=&\tau_{\rm in}(\ve,T)\,
{\cal E}^2_\omega\,\frac{\sigma^{\rm D}_\omega}{2\omega^2\nu_0}\\
&\times&\sum\limits_{\pm}\tilde{\nu}(\ve\pm\omega)
\,[\,f_T(\ve\pm\omega)-f_T(\ve)\,]. 
\label{f<}\eea
Equation (\ref{photo}) then reproduces Eq.~(\ref{linear}) for the
photoconductivity, with $\tau_{\rm in}$ (entering ${\cal P}_\omega$) given by
\be
\tau_{\rm in}=\int\! \frac{d\ve}{2\omega}\tau_{\rm
  in}(\ve,T)[f_T(\ve-\omega)-f_T(\ve+\omega)]. 
\label{tau_gen}
\ee
For $T\gg\omega$ this expression reduces back to Eq.~(\ref{tau>}), 
while in the opposite limit $T\ll\omega$ it yields
\be
\tau_{\rm in}=\int\! \frac{d\ve}{2\omega}\tau_{\rm in}(\ve,T)
=\frac{\pi^2 T}{2\omega}\,\tau_{\rm in}(0,T).
\label{tau_<}
\ee
Thus, the $T^{-2}$ scaling of the photoresistivity at $T\gg\omega$ transforms
into the $T^{-1}$ behavior for $T\ll\omega$.

\subsection{Separated Landau levels}
\label{ss62}

We turn now to the case of separated LLs. Let us assume again that $T$ is not
too high,
\be \omega_c\ll T
\ll\Gamma(\tau_{\rm tr}/\tau_{\rm q})^{1/2}~.
\ee
The dominant contribution, similarly to the case of overlapping LLs, is given
by large transferred momenta $q\gg (\Gamma/D_{\rm B})^{1/2}$, where $D_{\rm
B}=R_c^2/2\tau_{\rm tr,\,B}$ is the diffusion coefficient in a quantizing
magnetic field (see Sec.~\ref{s1}).  However, the situation for a strongly
oscillating DOS, Eq.~(\ref{SepDOS}), is different in that it is no longer
sufficient to deal solely with the out-scattering processes even for large
$q$; that is the whole collision integral (\ref{St}) should be taken into
account.  The kernel in Eq.~(\ref{St}) may be re-written as:
\bea
\label{A_SepLL}
A(E,\,&\ve,&\,\ve^\prime)\nonumber \\
&=&\frac{2}{\pi^3\nu_0\tilde\nu(\ve)}\int\!\frac{d^2 q}{(2\pi)^2}\,\,
|U|^2\,\Pi^q_{\ve,\,\ve+E}\,\Pi^q_{\ve',\,\ve'-E}\,,
\eea
where the function $\Pi$ for large $q\gg(\Gamma/D_{\rm B})^{1/2}$
reads\cite{khaetskii}
\bea
\Pi^q_{\ve,\ve'}=\pi\nu_0(q v_F)^{-1}\tilde\nu(\ve)\tilde\nu(\ve^\prime)\,.
\eea 
The procedure leading to Eq.~(\ref{St_tr}) is applicable in the case of
separated LLs as well. As a result, the inelastic collision term on the
right-hand side of Eq.~(\ref{kineq}) should be replaced by
\bea\label{coll}\nonumber
&&-{\rm St}_{\rm in}\{f\}=\frac{\pi^2
T^2+\ve^2}{4\pi\ve_F\tilde\nu(\ve)}\,\ln\frac{\kappa v_F\tau_{\rm
q}^{1/2}}{\omega_c\tau_{\rm tr}^{1/2}}\:\frac{\partial f_T}{\partial
\ve}\\\nonumber 
&&\times\,\langle\;\tilde\nu(\ve)\tilde\nu(\ve+E)
\tilde\nu(\ve^\prime)\tilde\nu(\ve^\prime-E)\\
&&\times\,[\,\varphi(\ve)-\varphi(\ve+E)
+\varphi(\ve^\prime)-\varphi(\ve^\prime-E)\,]
\;\rangle_{\ve^\prime,E}~.
\eea
In other words, Eq.~(\ref{kineq}) becomes an integral equation for the
periodic function $\varphi(\ve)$ characterizing oscillations of the
distribution function,
\bea\label{kineqSep}\nonumber
&&{\cal A}\,[\,\tilde{\nu}(\ve+\omega)-\tilde{\nu}(\ve-\omega)\,]
=\langle\;\tilde\nu(\ve+E)\,
\tilde\nu(\ve^\prime)\,\tilde\nu(\ve^\prime-E)\\
&&\times\,[\,\varphi(\ve)-\varphi(\ve+E)
+\varphi(\ve^\prime)-\varphi(\ve^\prime-E)\,]
\;\rangle_{\ve^\prime,E}~,
\eea
where ${\cal A}$ is a smooth function of $\ve$,
$$
{\cal A}=\frac{2\pi\ve_F{\cal E}^2_\omega\,\sigma^{\rm D}_\omega}
{\omega\nu_0(\pi^2 T^2+\ve^2)}\,
\ln^{-1}\frac{\kappa v_F\tau_{\rm q}^{1/2}}
{\omega_c\tau_{\rm tr}^{1/2}}\,.
$$

Analytical solution of Eq.~(\ref{kineqSep}) does not seem feasible.  However,
up to a factor of order unity,\cite{unity} we can rewrite the exact collision
integral (\ref{coll}) in the relaxation-time approximation, thus returning to
Eq.~(\ref{kineq}) with
\be
\label{sepLL} \tau_{\rm in}^{-1}\sim\frac{\omega_c}{\Gamma}\:\frac{
T^2+(\ve/\pi)^2}{\ve_F} \,\ln\frac{\kappa v_F\tau_{\rm
q}^{1/2}}{\omega_c\tau_{\rm tr}^{1/2}}~.
\ee
One sees that the $T$ and $\ve$ dependence of $\tau_{\rm in}$ in the regime of
separated LLs is the same as for overlapping LLs
[Eqs.~(\ref{out}),(\ref{tau_in})]. Therefore the temperature scaling of the
linear-in-${\cal P}_\omega$ photoresistivity, Eq.~(\ref{1order}), is the same
as found in Sec.~\ref{ss61}. In particular, $\sigma_{\rm ph}-\sigma_{\rm dc}$
scales as $T^{-2}$ for $T\gg\omega$.

\section{Oscillatory Hall resistivity}
\label{s7}

Finally, let us briefly discuss the issue of the microwave induced
$\omega/\omega_c$--oscillations of the Hall resistivity $\rho^H_{xy}$
(antisymmetric part of $\rho_{xy}$, i.e. $\rho^H_{xy}=-\rho^H_{yx}$) detected
in recent experiments.\cite{Rxy1} The experimentally observed oscillations of
$\rho_{xy}$ demonstrate the following properties. First, they have the same
period and the opposite phase as compared to the oscillations of the
dissipative resistivity $\rho_{xx}$. Second, the amplitude of the oscillations
in $\rho_{xy}$ is roughly the same as in $\rho_{xx}$. Third, the
radiation-induced contribution $\delta\rho_{xy}$ is odd with respect to the
magnetic field $B$, which implies that the measured $\delta\rho_{xy}$ is
indeed a contribution to the Hall resistivity $\rho^H_{xy}$.  These
observations cannot be explained within either the mechanism related to the
effect of the microwaves on the distribution function
(Ref.~\onlinecite{dmitriev03} and this work) or the one related to the effect
on the elastic collision integral
(Refs.~\onlinecite{durst03,ryzhii,vavilov03}), if one assumes, as usual, that
the electron density $n_e$ is constant. If $n_e$ was constant, the leading
odd-in-$B$ correction $\delta\sigma^H_{yx}$ to the Hall conductivity
$\sigma^H_{yx}=ce n_{\rm e}/B$ should be smaller than $\delta\sigma_{xx}$ by a
factor $1/\omega_c\tau_{\rm tr}$, which is of order $10^{-2}$ under the
experimental conditions. Therefore, the observed $\delta\rho^H_{xy}$, although
very small in comparison with the bare $\rho_{xy}$, appears to be two order of
magnitude larger than what one would expect from the theory of the oscillatory
contribution to $\rho_{xy}$ neglecting the oscillations of
$n_e$.\cite{note-ryzhii} A possible origin of the observed oscillatory
$\rho_{xy}$ may be a weak variation of $n_e$ with $\omega/\omega_c$. The
observation\cite{Rxy2} of a variation of the period of the Shubnikov-de Haas
oscillations, $\delta\sigma_{xx}\propto \cos (c\,n_{\rm e}/eB)$, appears to
support the idea of the microwave-induced oscillations of the electron
density. The issue warrants further study.

\section{Conclusion}
\label{s8}

To summarize, we have presented a theory of magnetooscillations in the
photoconductivity of a 2DEG. The parametrically largest contribution to the
effect is governed by the microwave-induced change in the distribution
function. We have analyzed the nonlinearity with respect to both the microwave
and {\it dc} fields. The result takes an especially simple form in the regime
of overlapping LLs, Eq.~(\ref{result}). We have shown that the magnitude of
the effect is governed by the inelastic relaxation time (\ref{tau_gen}),
(\ref{tau_in}), (\ref{sepLL}), and increases as $T^{-2}$ or $T^{-1}$
(depending on the relation between $T$ and $\omega$) with lowering
temperature.  For a sufficiently strong microwave power the linear-in-{\it
dc}-field photoconductivity becomes negative leading to formation of domains
with zero resistivity. We have calculated the threshold power at which this
zero--resistance state is formed, Eq.~(\ref{P*}), and the spontaneous {\it dc}
field in the domains, Eq.~(\ref{e-domain}).

Our results are in overall agreement with the experimental findings.
\cite{mani02,zudov03,note-ryzhii2} The observed $T$ dependence of the
photoresistivity at maxima compares well with the predicted $T^{-2}$
behavior. Typical parameters $\omega/2\pi\simeq 50-100$~GHz,
$\tau_{\rm q}\simeq 10$~ps yield $\omega\tau_{\rm q}/2\pi\simeq 0.5-1$
(overlapping LLs), and the experimental data indeed closely resemble
Fig.~\ref{fig1}. For $T\sim 1\:{\rm K}$ and $\epsilon_F\sim 100\:{\rm
K}$ we find $\tau_{\rm in}^{-1}\sim 10\:{\rm mK}$, much less than
$\tau_{\rm q}^{-1}\sim 1\:{\rm K}$, as assumed in our theory. For the
microwave power $P\sim 1$~mW and the sample area $S\sim 1\,{\rm
cm}^2$, Eq.~(\ref{P0}) yields the dimensionless power ${\cal
P}_\omega^{(0)}\sim 0.005-0.1$ (the smaller value corresponds to
$\omega/2\pi=100$~GHz, the larger one to $\omega/2\pi=50$~GHz), where
we used $\tau_{\rm tr}=10$~mK and $v_F=2\cdot10^7$~cm/s.  These values
of ${\cal P}_\omega^{(0)}$ agree with characteristic values for
separated LLs (Fig.~\ref{fig3}) but are noticeably less than the
prediction for overlapping LLs (Fig.~\ref{fig1}).  The discrepancy can
be attributed (at least partly) to the fact that the value of
$\tau_{\rm q}$ used in the above estimate, which was extracted from
the Shubnikov--de Haas experiments, is in fact masked by inhomogeneous
broadening and thus is shorter than the actual value.  Indeed,
$\tau_{\rm q}$ found from the (experimental) damping of the
oscillations in $\rho_{\rm ph}$ (which are local and thus not affected
by inhomogeneous broadening) according to Eq.~(\ref{result}) is
several times longer.  With this value of $\tau_{\rm q}$ the threshold
microwave power ${\cal P}_\omega^*$, Eq.~(\ref{P*}), needed for the
emergence of the zero-resistance states, corresponds to $P$ less than
$1\:{\rm mW}$, in conformity with the experiments.\cite{conformity} Finally, for ${\cal
P}_\omega-{\cal P}_\omega^*\sim{\cal P}_\omega^*$, at $T\sim1$~K (when
$\tau_{\rm tr}/\tau_{\rm in}\sim 1$), and $\omega_c/2\pi=50$~GHz the
estimated {\em dc} electric field in the domains,
Eq.~(\ref{e-domain}), is found to be ${\cal E}_{\rm dc}^*\sim 1\:{\rm
V/cm}$. This is in agreement with the experimental data of
Ref.~\onlinecite{willett03} where the voltage drop between an internal
and an external contact (separated by 200~${\rm \mu m}$) generated by
the radiation in the absence of the drive current was of the order of
5~mV for $\omega_c/2\pi\simeq20$~GHz. Assuming that the size of the
domain is of the order of the system size, this yields ${\cal E}_{\rm
dc}^*\sim 0.25\:{\rm V/cm}$. One sees that this value indeed compares
well with our theoretical estimate ${\cal E}_{\rm dc}^*\sim 1\:{\rm
V/cm}$, especially taking into account the $\omega_c^2$ dependence of
${\cal E}_{\rm dc}^*$ following from Eq.~(\ref{e-domain}).

Recently, a number of publications appeared that extended our
theory (main results of which were presented in Ref.~\onlinecite{short})
in a variety of contexts: propagation of surface-acoustic
waves,\cite{falkoprl,falkoc-m} photoconductivity of
laterally-modulated structures,\cite{vonoppen} photoconductivity for
$B$ above the cyclotron resonance,\cite{dorozhkinnew} local
compressibility of irradiated samples.\cite{compress}

We thank R.R.~Du, K.~von~Klitzing, R.G.~Mani, J.H.~Smet, and
M.A.~Zudov for information about the experiments, and I.V.~Gornyi for
numerous stimulating discussions.  This work was supported by the SPP
``Quanten-Hall-Systeme'' of the DFG, by NSF grants DMR02-37296, DMR
02-13282 and EIA02-10376, by AFOSR grant F49620-01-1-0457, and by the
RFBR.


\appendix
\section*{Derivation of the basic equations
from the quantum Boltzmann equation}
\label{app1}

The purpose of this Appendix is to demonstrate how the semiclassical transport
theory of Ref.~\onlinecite{vavilov03} (see Sec. III of this reference) is
reduced to Eqs.~(\ref{photo}),(\ref{dc_short}),(\ref{kineq1}) when the effect
of electric fields on the impurity collision process can be neglected and only
the distribution function is affected. The conditions under which the effect
of the fields on the collision integral is weak are that the $dc$ field is
much smaller than $E_0$ and the strength of the $ac$ field satisfies ${\cal P}
\ll 1$, where $E_0$ and ${\cal P}$ are defined in Eqs.~(5.5) and (6.2) of
Ref.~\onlinecite{vavilov03}, respectively. In notation of our
Eq.~(\ref{unitsPQ}) these conditions read\cite{bytheway}
\be
{\cal P}_\omega,\ {\cal Q}_{\rm dc} \ll \tau_{\rm in}/\tau_{\rm q}~.
\label{ap:cond}
\ee
Under the conditions (\ref{ap:cond}) we neglect the effect of the external
field on the density of states, which amounts to putting $h_1=1$ in Eq.~(3.42)
of Ref.~\onlinecite{vavilov03}.  The resulting DOS is time independent:
\be
\nu(\ve)  =\nu_0
\left[1+ 2 {\rm Re}\sum_{l=1}^{\infty} \lambda^l g_l \exp
\left( \frac{i 2\pi
    l\ve} 
{\omega_c}
\right)
\right],
\label{nue}
\ee
where the coherence factor $\lambda=-\exp(-\pi/\omega_c\tau_q)$,
\[
g_l  = l^{-1}\,L_{l-1}^1(2\pi l/\omega_c\tau_{\rm q}),
\]
and $L_l^m$ is the Laguerre polynomial.  The combination $|g_l\lambda^l|^2$
has the meaning of the probability for an electron to complete the cyclotron
orbit after $l$ revolutions.  Here we follow the notations of
Ref.~\onlinecite{vavilov03}; in the main text of the present paper the factor
$\lambda$ is denoted as $-\delta$.

The kinetic equations (3.46) of Ref.~\onlinecite{vavilov03} are written in the
time representation for the distribution function $f(t,t^\prime; \varphi,
\R)$, where $\phi$ is the angle on the cyclotron orbit and $\R$ is the
position of its guiding center.  This function is related to the conventional
distribution in the energy-time representation by the Wigner transform
\be
f(t,t^\prime)=\int\frac{d\ve}{2\pi}e^{-i\ve(t-t^\prime)}f\left(
\frac{t+t^\prime}{2},\ve\right).
\label{WT}
\ee
The inverse Wigner transform of both sides of Eq.~(3.46a) of
Ref.~\onlinecite{vavilov03} yields the canonical form of the Boltzmann
equation for the distribution function
\be
(\partial_t+ \w_c \partial_\varphi)
f\left(t,\ve; \varphi,\R\right)
= \St_{\rm im}\{f\}_{\ve, t} + \St_{\rm in}\{f\}_{\ve,t}.
\label{Boltzmann}
\ee
A time independent version $\St_{\rm in}\{f\}_{\ve,t}$ of the inelastic term
is considered in Sec.~\ref{s6} and will not be discussed further in this
Appendix.  The impurity collision term $\St_{\rm im}\{f\}_{\ve, t}$ is
obtained as the inverse Wigner transform of Eq.(3.46b) of
Ref.~\onlinecite{vavilov03} and will be written explicitly below.

Under the condition (\ref{ap:cond}), the dissipative electric current at the
point $\r$ is obtained from Eq.~(\ref{WT}) and Eq.~(3.45b) of
Ref.~\onlinecite{vavilov03}:
\bea\label{ap:current}
&&\j^{(d)}(\r,t)=2 e p_F\int\limits_0^{2\pi}
\frac{d\varphi}{2\pi}\nn(\phi)
\int \frac{d\ve}{2\pi}
\Bigg\{f\left(t,\ve; \varphi,\r_g
\right)
\nonumber
\\
&&+  2{\rm Re} \sum_{l=1}^{\infty}\lambda^l
g_{l}\exp\left(\frac{i2\pi\ve}{\w_c}\right)
 f\left(t-\frac{\pi l}{\w_c}, \ve; \varphi,\r_g\right)
\Bigg\},
\eea
where $p_F$ is the Fermi momentum,
$\r_g=\r-R_c\,\hat\epsilon\,\nn(\varphi)-\bzeta(t)$ is the guiding center
coordinate and $\bzeta(t)$ describes the motion of an electron in the external
field $\E(t)$:
\be
\partial_t \bzeta(t)=
\left(\frac{\partial_t -\w_c \hat{\epsilon}}{\partial_t^2+\w_c^2}\right)
\frac{e\E(t)}{m},
\label{zeta}
\ee
$\nn=\{\cos\varphi,\sin\varphi\}$, 
$\hat{\epsilon}$ is the antisymmetric tensor,
$\epsilon_{xy}=-\epsilon_{yx}=1$.

In view of Eq.~(\ref{ap:cond}) one can neglect the effect of the electric
fields on the interference processes in the collision integral (3.46b) of
Ref.~\onlinecite{vavilov03}, which amounts to putting the form-factors
$h_1=h_2=1$ in Eq.~(3.49).  Then Eq.~(3.46b) in the time representation
acquires a simple form
\bea
&& \St_{\rm im}\{f\}_{t, t^\prime}= 
-\frac{\hat{L}(t,t^\prime)}{\tau_{\rm tr}}
f(t,t^\prime) 
\label{collision}
\\
&+& \frac{1}{\tau_{\rm tr}}
\sum_{l=1}^\infty\lambda^lg_l\left\{
 \left[ M_l(t)-  \hat{L}\left(t,t^\prime \right)
\right] f\left(t-\frac{2\pi l}{\w_c},t'\right) 
\right\}
\nonumber
\\
&+& \frac{1}{\tau_{\rm tr}}
\sum_{l=1}^\infty\lambda^lg_l\left\{
 \left[ M_l(t^\prime)-  \hat{L}\left(t,t^\prime \right)
\right] f\left(t,t'-\frac{2\pi l}{\w_c}\right) 
\right\},
\nonumber
\eea
where
$$
\hat{L}(t,t^\prime) = \left(\,\{\,p_F[\,\bzeta(t)- \bzeta(t^\prime)\,]
  \hat\epsilon -
i R_c\nabla_R\,\}\cdot \nn(\varphi)+ i\partial_\varphi\,\right)^2
$$
and $M_l(t)$ is the result of action of the operator $\hat{L} \left(t,\,t-2\pi
l/\w_c\right)$ on unity.  Performing the Wigner transformation of
Eq.~(\ref{collision}) we obtain
\bea
 &&\St_{\rm im}\{f\}_{\ve,t}= \nonumber \\
&-&\frac{1}{\tau_{\rm tr}}\hat{L}\left(t+\frac{i}{2}\partial_\ve\,,\,
t-\frac{i}{2}\partial_\ve
\right)
f(t,\ve;\varphi,\R)
+ \frac{2}{\tau_{\rm tr}}{\rm Re}
\sum_{l=1}^\infty\lambda^lg_l
\nonumber\\
&\times& \Bigg\{
 \left[ M_l\left(t+\frac{i}{2}\partial_\ve\right)
-  \hat{L}\left( t+\frac{i}{2}\partial_\ve\,,\,
t-\frac{i}{2}\partial_\ve\right)
\right]
\nonumber\\
&\times&
 \exp\left({\frac{2i \pi l\ve}{\w_c}}\right)\,
f\left(t-\frac{\pi l}{\w_c},\ve;\varphi,\R\right) 
\Bigg\}.
\label{collisionWT}
\eea

Equations (\ref{Boltzmann}) and (\ref{collisionWT}) are valid for an arbitrary
time dependent distribution function.  We now notice from
Eq.~(\ref{Boltzmann}) that the most divergent terms in the distribution
function are due to the contribution of the angular and time independent part
of $f$ to the collision integral.  For $\w_c\tau_{\rm tr}\gg 1$ it is
sufficient to consider $f$ to be time, angular, and coordinate
independent. Equation (\ref{collisionWT}) reduces to
\bea
&&\overline{\St_{\rm im}\{f\}_{\ve,t}}= 
-\frac{
  N\left( i\partial_\ve\right)
}{\tau_{\rm tr}}
f(\ve)
\nonumber
+ \frac{2}{\tau_{\rm tr}}{\rm Re}
\sum_{l=1}^\infty\lambda^lg_l
 \\\nonumber&\times&\Bigg\{
 \left[\,  N\left(\frac{2\pi l}{\w_c}\right)
-  N\left( i\partial_\ve\right)
\,\right]
 \exp \left(\frac{2i \pi l\ve}{\w_c}\right)
f\left(\ve\right) 
\Bigg\},
\nonumber
\\
 &&N\left(\delta t\right) \equiv  \frac{p_F^2}{2}
\overline{\left[\,\bzeta(t)- 
\bzeta\left(t-\delta t\right)
\,\right]^2}.
\label{collisionWT2}
\eea  
and the bar stands for the time averaging.  In the constant electric field
$N\left(\delta t\right)=(\delta t)^2 p_F^2 (\partial_t\bzeta)^2/2$, whereas
for the microwave field $\bzeta(t) ={\rm Re}\, (\bzeta_\w e^{i\w t})$ and
$N\left(\delta t\right)= p_F^2 {\rm Re}\,
[\,\bzeta_\w\bzeta_\w^*(1-e^{i\w\delta t})\,]/2$.  Substituting these
expressions in Eq.~(\ref{collisionWT2}), taking $\bzeta$ from
Eq.~(\ref{zeta}), and using Eq.~(\ref{nue}), we arrive at the left-hand side
of Eq.~(\ref{kineq}).

The zero angular harmonic of the distribution function does not contribute to
the electric current (\ref{ap:current}) directly.  The relevant angular
dependent correction $\delta f(\varphi)$, with $\int d\varphi\,\delta
f(\varphi)=0$, can be found perturbatively from Eqs.~(\ref{Boltzmann}) and
(\ref{collisionWT}):
\bea
&& \w_c \partial_\varphi
 \delta f\left(\varphi\right) \label{deltaf1}
 =p_F\tau_{\rm tr}^{-1}\partial_\varphi
 \left[\,\overline{\partial_t\bzeta(t)} \,\hat\epsilon\, \nn(\varphi) \,\right]
\nonumber
\\
&& \times 
 \left[\,1 + 2 {\rm Re}\sum_{l=1}^\infty\lambda^lg_l 
\exp \left(\frac{2\pi
   i l \ve}{\w_c}\right)\,\right]
 \partial_\ve f(\ve).
\label{deltaf}
 \eea
Substituting the solution of Eq.~(\ref{deltaf1}) in Eq.~(\ref{ap:current}) and
applying Eq.~(\ref{zeta}) in the limit of the $dc$ field, we obtain
Eq.~(\ref{photo}) with $\sigma_{\rm dc}(\ve)$ given by Eq.~(\ref{dc_short}).


\end{multicols}

\begin{references}
\bibitem{note3} As a matter of fact, the analytic results work
numerically well down to $T\sim\omega_c$. The reason is that the
actual small parameter is $\omega_c/2\pi T\ll 1$. In particular, this
is illustrated in Fig.~\ref{fig0}, where $T=\omega_c$ but the scale
separation between the fast oscillations and the smooth energy
dependence of $f_T(\ve)$ is evident. Related to that is the
temperature damping of the Shubnikov-de Haas oscillations
$\exp(-2\pi^2 T/\omega_c)$.
\bibitem{note4} Strictly speaking, Eq.(\ref{distr}) is written under
the assumption $T\gg\omega$ which is stronger than the condition $T\gg
\omega_c$ for high oscillation harmonics, $\omega/\omega_c\gg 1$. This
condition, however, only helps to simplify technically the solution of
Eq.(\ref{kineq1}) but is not essential for the physics, which is based
on the separation of energy scales (fast oscillations vs. smooth
dependence) and requires $T\gg\omega_c$ only. In particular, it is not
difficult to check (see Sec.~\ref{s6}) that the
linear-in-microwave-power result (\ref{linear}) remains valid
independently of the relation between $T$ and $\omega$.
\bibitem{note1} One can show that for an arbitrary $\nu(\ve)$,
 Eq.~(\ref{kineq1}) yields for ${Q}_{\rm dc} \gg 1$ the conductivity
 $\sigma=\sigma_{\rm dc}^{\rm
 D}/\langle\tilde{\nu}^{-2}(\ve)\rangle_{\ve}$, which is positive for
 overlapping LLs and vanishes in the presence of spectral gaps.
\bibitem{crossover} For $\Delta_n\sim 1$, $A_n^<(E)$ is governed by
momenta $q\sim q_1$, so that $A_n^<(E)\sim1/\ve_F$.
\bibitem{note2} The energy $\ve$ is counted in Sec.~\ref{s6}
from the Fermi level.
\bibitem{conserve} Note that despite the energy dependence of
$\tau_{\rm in}$ on the scale of $\ve\sim T$, the collision integral in
(\ref{tau_in}) conserves the number of particles in view of fast
oscillations of $f-f_T$ with a period $\omega_c\ll T$.
\bibitem{unity} Strictly speaking, this factor depends on the argument
$(\omega-n\omega_c)/\Gamma$ of the function $\Phi$ in
Eq.~(\ref{1order}), which describes the shape of oscillations in
Fig.~\ref{fig3}.  Therefore, the precise shape of oscillations will be
somewhat modified if one uses the exact collision integral
(\ref{coll}) instead of the relaxation time approximation
(\ref{kineq}).  However, all the qualitative features of
Fig.~\ref{fig3} will remain intact.
\bibitem{bytheway} Throughout the paper we assume that $\tau_{\rm
in}/\tau_{\rm q}\gg 1$. Under this condition, the neglected effect of
the electric fields on the collision integral is much weaker than the
effect on the distribution function, retained in Eq.~(\ref{kineq}).
\bibitem{note-ryzhii2} It was claimed in recent papers
\cite{ryzhii03,ryzhii04} that the effect of microwaves on the
collision integral studied in
Refs.~\onlinecite{durst03,ryzhii,vavilov03} can explain the
experimental data. We disagree with these claims. First, this effect
is small compared to the one studied in the present paper by a factor
$\tau_q/\tau_{\rm in}$ (see Sec.~\ref{s3}), which is of order
$10^{-2}$ at $T\sim 1\:{\rm K}$, and is thus much smaller than what is
experimentally observed (unfortunately, the results of
Refs.~\onlinecite{ryzhii03,ryzhii04} are presented in the form of
proportionality relations, which makes a comparison of the magnitude
of the effect to the experiment impossible). An agreement with the
experimental data claimed in Ref.\onlinecite{ryzhii03} may be due to
the fact that the authors used for their estimate the microwave power
provided by a Gunn diode,\cite{zudov03} while the attenuated power
incident on the sample was smaller by more than two orders of
magnitude. Second, the contribution related to the collision integral
is $T$-independent and polarization-dependent (as discussed in
Sec.~\ref{s3}), in contrast to the experiment.  An
attempt\cite{ryzhii04} to explain the temperature dependence by a
$T$-dependent LL width determined by the electron-electron interaction
is in contradiction with the fact that the inelastic scattering rate
$\tau_{\rm ee}^{-1}$ is two orders of magnitude smaller than the
impurity-scattering rate $\tau_q^{-1}$ at $T\sim 1\:{\rm K}$.
\bibitem{conformity} Interpolating between Figs.~\ref{fig1} and
  \ref{fig3}, we estimate the dimensionless microwave power necessary
  for the zero-resistance state according to our theory as ${\cal
    P}^{(0)}_\omega\sim$~0.05. This value agrees with the
  dimensionless power necessary for 50~GHz (${\cal
    P}^{(0)}_\omega=0.1$), which we estimate using the experimental
  data [according to Eq.~(\ref{P0})] and assuming that there are no
  standing waves in the cavity. There remains, however, some
  discrepancy with the value of power (0.005) for 100~GHz, which might
  possibly be attributed to the fact that both the power and the
  distribution of the electromagnetic field in the cavity are not very
  well known in the current experiments.
\end{references}
\end{document}